\documentclass[a4paper]{JHEP3}
\usepackage[english]{babel}
\usepackage{amsmath,amssymb}
\usepackage[dvips]{graphicx}
\usepackage{amsfonts}
\usepackage{bbm}
\usepackage{cite}
\newcommand{\bc}{\begin{center}}
\newcommand{\ec}{\end{center}}
\newcommand{\bt}{\begin{tabular}}
\newcommand{\et}{\end{tabular}}
\newcommand{\be}{\begin{equation}}
\newcommand{\ee}{\end{equation}}
\newcommand{\bea}{\begin{eqnarray}}
\newcommand{\eea}{\end{eqnarray}}
\newcommand{\bfig}{\begin{figure}}
\newcommand{\efig}{\end{figure}}
\newcommand{\ba}{\begin{array}}
\newcommand{\ea}{\end{array}}
\newcommand{\bi}{\begin{itemize}}
\newcommand{\ei}{\end{itemize}}
\newcommand{\ie}{{\it i.e. }}
\newcommand{\bfr}{\begin{flushright}}
\newcommand{\efr}{\end{flushright}}
\newcommand{\bfl}{\begin{flushleft}}
\newcommand{\efl}{\end{flushleft}}
\def\call{{\cal L}}
\def\calu{{\cal U}}
\def\e{{\rm e}}
\preprint{ULB-TH/10-20 \\
INR-TH-2010-37}
\title{See-saw neutrino masses and large mixing angles \\
in the vortex background on a sphere}
\author{J.-M.~Fr\`ere$^{\hbox{a}}$, M. Libanov$^{\hbox{b}}$, F.-S. Ling$^{\hbox{a}}$\\
$^{\hbox{a}}$Service de Physique Th\'eorique, Universit\'e Libre de Bruxelles, \\
Campus de la Plaine CP225, Bd du Triomphe, 1050 Brussels, Belgium\\
$^{\hbox{b}}$Institute for Nuclear Research of the Russian Academy of Sciences,\\
60th October Anniversary Prospect 7a, 117312, Moscow, Russia\\
E-mail: \email{frere@ulb.ac.be}; \email{ml@ms2.inr.ac.ru}; \email{fling@ulb.ac.be}}
\abstract{In the vortex background on a sphere, a single 6-dimensional fermion family gives rise to 3 zero-modes
in the 4-dimensional point of view, which may explain the replication of families in the Standard Model.
Previously, it had been shown that realistic hierarchical mass and mixing patterns can be reproduced for
the quarks and the charged leptons.
Here, we show that the addition of a single heavy 6-dimensional field that is gauge singlet,
unbound to the vortex, and embedded with a bulk Majorana mass
enables to generate 4D Majorana masses for the
light neutrinos through the see-saw mechanism.
The scheme is very predictive. The hierarchical structure of the fermion zero-modes leads
automatically to an inverted pseudo-Dirac mass pattern, and always predicts one maximal angle in the
neutrino see-saw matrix. It is possible to obtain a second large mixing angle from either the charged lepton or
the neutrino sector, and we demonstrate that this model can fit all observed data in neutrino oscillations experiments.
Also, $U_{e3}$ is found to be of the order $\delta \sim 0.1$.}


\begin{document}
%
\section{Introduction}
\label{sec:intro}

Models of particle physics in more than four spacetime dimensions offer interesting possibilities
to explain the mysterious patterns observed in fermion intra- and inter-family mass hierarchies and mixings.
Standard Model fields are localized near a four dimensional subspace, in the core of a topological
defect in extra dimensions.
With different wave function profiles for different fermions, their overlap with a scalar field $H$ that
plays the role of the Standard Model Brout-Englert-Higgs field (BEH scalar) leads to a
hierarchical structure of masses and mixings from the four dimensional point of view.

The class of models where the topological defect is a two dimensional Abelian vortex, known as
the Abrikosov-Nielsen-Olesen vortex is of particular interest as it leads to chiral fermion zero modes.
Through the index theorem, the number of families in the Standard Model, $n_f=3$, can be related to
the number of zero modes as a topological invariant of the background vortex field.
The value $n_f=3$ is however not automatically guaranteed, but can be achieved by an adequate axial charge
assignment for the fermions.
In Refs.~\cite{Libanov:2000uf, Frere:2000dc, Frere:2003yv, Libanov:2002tm,
Libanov:2002ka}, it has been demonstrated that the
different wave function profiles of the zero modes in the core of the
vortex lead naturally to realistic hierarchical mass patterns for quarks
and charged leptons, with small inter-generation mixings. The model was
first developed assuming two flat extra dimensions $R^2$, but later it
appeared that a compactification on a sphere $S^2$ is necessary for the
model to be consistent and realistic. The phenomenological implications
also include a specific pattern of flavor violation where ``family number"
non conserving processes are automatically strongly
suppressed~\cite{Frere:2003ye}. Such FCNC effects mediated by heavy bosons
can be searched for at LHC and future colliders, with the production of
$(\mu^+ e^-)$ or $(\mu^- \tau^+)$ pairs with equal and large transverse
momenta as the main signature~\cite{Frere:2004yu}.

In the current paper, we want to consider whether the scheme can be extended to accommodate mass and mixing data
in the neutrino sector. Neutrinos are not only characterized by tiny masses, at most in the eV range,
but also exhibit large mixings responsible for the observed neutrino flavor oscillations.
An obvious possibility to generate neutrino masses would be to treat them exactly like the charged fermions,
with a Dirac mass obtained at the cost of introducing a 6D field $N$, bound to the vortex,
and from which the three families of 4D right-handed neutrinos emerge.
However, this possibility does not offer a natural explanation for the smallness of the neutrino masses,
which in this case require tiny coefficients in the Lagrangian.

It is therefore tempting to consider other solutions, namely the case where the ``right-handed"
neutrino field is NOT bound to the vortex.
In the context of models with large extra dimensions, tiny neutrino masses are often the result of
a dilution effect: the field that provides right-handed neutrinos, being singlet under
the Standard Model gauge group, can be non-localized, and therefore have a small overlap with the wave function
of Standard Model fields\footnote{In 5D, tiny neutrino masses as the result of multi-localization
of wave-functions have been considered in Ref.~\cite{Frere:2003hn} for flat extra-dimensions, and
in Refs.~\cite{Chang:2005ya, Moreau:2005kz} for warped extra-dimensions.}.
A first attempt using a non-localized field $N$ was made in
Ref.~\cite{Frere:2001ug}, in a setup where
the two extra dimensions were flat.
When the size $R$ of the compact extra dimensions is small enough to neglect the massive modes of the
Kaluza-Klein tower (which is necessary to avoid astrophysical bounds on neutrinos derived from
supernova dynamics), we found that the 6D field reduces in the 4D effective theory to two right-handed
and two left-handed states, the former being suitable for building Dirac mass terms for the neutrinos.
These mass terms  are furthermore reduced according to the limited overlap of the $L$ and $R$ fields,
respectively bound and unbound to the vortex.
With two right-handed neutrinos, the model is able to accommodate the two mass
squared differences measured in solar and atmospheric neutrinos oscillation experiments.
However, after masses of $e$, $\mu$ and $\tau$ are fixed,
the model is essentially parameter free and therefore also predicts the neutrino mixing angles to be
$\sin \theta_{12} \simeq 0.37$, $\sin \theta_{13} \simeq 0.14$, $\sin \theta_{23} \simeq 0.99$,
a pattern that differs significantly from the observed one.

In a three neutrino scheme, the observed mixing matrix in the leptonic sector,
the so-called Maki-Nakagawa-Sakata (MNS) matrix which relates neutrino flavor eigenstates
to mass eigenstates, $\nu_f = U_{MNS} \; \nu_m$, is conventionally parametrized
in the following form~\cite{Amsler:2008zzb}
\be
U_{MNS} = \left(
\ba{ccc}
    c_{12} c_{13}
    & s_{12} c_{13}
    & s_{13} e^{-i\delta_\text{CP}}
    \\
    - s_{12} c_{23} - c_{12} s_{13} s_{23} e^{i\delta_\text{CP}}
    & \hphantom{+} c_{12} c_{23} - s_{12} s_{13} s_{23} e^{i\delta_\text{CP}}
    & c_{13} s_{23} \hspace*{5.5mm}
    \\
    \hphantom{+} s_{12} s_{23} - c_{12} s_{13} c_{23} e^{i\delta_\text{CP}}
    & - c_{12} s_{23} - s_{12} s_{13} c_{23} e^{i\delta_\text{CP}}
    & c_{13} c_{23} \hspace*{5.5mm}
\ea
\right) \quad ,
\label{eq:umns}
\ee
where $s_{ij} = \sin \theta_{ij}$, $c_{ij} = \cos \theta_{ij}$, $\delta_\text{CP}$ is a CP violating phase,
and where two additional Majorana phases which do not affect oscillations have been dropped.
The latest update on a three neutrino global fit gives
$\tan^2 \theta_{12} = 0.47_{-0.10}^{+0.14}$, $\tan^2 \theta_{23} = 0.9^{+1.0}_{-0.4}$,
$\sin^2 \theta_{13} \leq 0.05$ at $3\sigma$ C.L.
The $CP$ phase is left unconstrained, and the mass squared splittings $\Delta m_{ij}^2 = m^2_i-m^2_j$
are determined as $\Delta m^2_{21} = 7.6 \pm 0.7 \times 10^{-5}~{\rm eV^2}$,
$\Delta m^2_{31} = 2.46 \pm 0.37 \times 10^{-3}~{\rm eV^2}$ (normal scheme) or
$\Delta m^2_{31} = -2.36 \pm 0.37 \times 10^{-3}~{\rm eV^2}$ (inverted
scheme)~\cite{GonzalezGarcia:2010er}.
Unlike the CKM matrix in the quark sector, the MNS matrix harbors two large mixing angles
seen in solar and atmospheric neutrino oscillation experiments,
together with a rather mild mass hierarchy, if any.

We have meanwhile turned to a different compactification scheme (namely,
on a sphere of radius $R$)~\cite{Frere:2003yv}. This compactification
scheme offers two advantages. It removes both the need of a difficult
localization of the gauge fields, and also spurious fermionic
zero-modes which may appear due to a boundary in extra dimensions.

It is thus now appropriate to reconsider the situation in the new context.
In particular, we want to investigate how it is possible to generate a pattern where the three light
neutrinos in fact behave like Majorana particles in four dimensions
(although Majorana fermions don't exist in 6D (see \emph{e.g.}
Ref.~\cite{Pilaftsis:1999jk}),
what we discuss here is the effective theory after dimensional reduction)\footnote{
Majorana neutrinos and the see-saw mechanism in 5D have been studied in 
Refs.~\cite{Huber:2003sf, Chen:2005mz, Haba:2006gt, Csaki:2008qq, 
Brakke:2008ka, Perez:2008ee, Watanabe:2009br, Chen:2009gy, Watanabe:2010cy, Kadosh:2010rm, Blennow:2010zu} 
for both flat and warped geometries. Usually the main problem is to explain the presence of 
the two large mixing angles. One solution that is advocated in the literature is by invoking a suitable
discrete symmetry.}.
For this purpose, with a ``see-saw" mechanism in mind , we add one extra (sterile) ``neutrino" field $N$
in 6D, unbound to the vortex. For this sterile particle (often called right-handed
neutrino), we expect two sources of mass: one arising
from the compactification itself (there are no massless modes in this
case due to the positive curvature of a sphere~\cite{Witten:1983ux,
Bailin:1987jd})\footnote{Note however that massless modes (without a
Majorana mass term)
can appear in other compactification schemes, with
flat or negative curvature. Therefore, the essential ingredient to implement the see-saw mechanism
is still the presence of a Majorana mass term for $N$, like in four dimensional theories.},
and the other, optional, from possible Majorana mass terms.

As a matter of nomenclature, it is indeed important to keep in mind
that, although Majorana fermions \ie self charge conjugated particles
don't exist in 6D, nothing prevents Majorana mass terms. By the latter, we
mean a scalar term in the Lagrangian, which violates fermion number. Thus
two effects can concur to lower the observed neutrino mass: the see-saw
mechanism, associated to the coupling to a heavy right-handed neutrino
(where the mass results either from compactification or from a Majorana
mass term), and the limited overlap between the unbound neutrino field and
the vortex-bound left-handed neutrino.
We anticipate here on the following to announce that in fact, the Majorana
mass term will prove necessary (see Sec.~\ref{sec:M0}) even in the presence of other
fermion number violating couplings.

The paper is organized as follows.
In Sec.~\ref{sec:majorana}, we first discuss the meaning and the properties of a Majorana mass term in 6D;
then, we give the decomposition of a 6D singlet neutrino field $N$ into spherical modes,
and reduce its Lagrangian including a bulk Majorana mass term  to a 4D effective Lagrangian.
In Sec.~\ref{sec:see-saw}, the field $N$ is coupled to the vortex-bound lepton doublet $L$ and see-saw masses
for the left-handed neutrinos are obtained.
We consider two possibilities: when the bulk Majorana mass $M$ is absent, explicit fermion number violating
couplings between $L$ and $N$ are introduced, when $M \neq 0$, the Majorana mass term is the only source
of fermion number violation.
Within the same framework, charged lepton masses are also calculated.
In Sec.~\ref{sec:matrices}, we estimate the mass matrices in the narrow BEH scalar approximation, and compare
them with phenomenological data. In particular, we discuss how the model can accommodate at the same time
hierarchical masses for the charged leptons, sub-eV neutrino masses with a mild hierarchy in their $\Delta m^2$,
and large angles seen in the mixing matrix $U_{MNS}$.
Finally, we conclude in Sec.~\ref{sec:end}.
Also, notations and definitions of $\Gamma$ matrices in 6D are summarized in the Appendix.

\section{Majorana mass term in six dimensions}
\label{sec:majorana}

\subsection{General properties}

As already stated, we use the expression ``Majorana mass term" for
a 6D Lorentz scalar contribution in the Lagrangian, typically breaking
fermion number by 2 units (assuming that fermion number has been previously defined).
All such terms can be constructed from the component expansion of the spinors.
We have explicitly checked  that all such bilinear contributions can also be
generated from the ``Dirac" spinors using one suitable ``charge conjugation matrix" $C$.
At least two definitions of the matrix $C$ exist, but taking  $C = \Gamma_0 \Gamma_2 \Gamma_4$
(up to a phase), the two possible Majorana mass terms that can \emph{a priori} be constructed
with two fields $\Psi$ and $\Phi$ (assuming them to have equal fermion number) are
\bea
&(S)&	 \quad  \bar{\Psi}^c \Phi + {\rm h.c.} \nonumber \\
&(A)&	 \quad  \bar{\Psi}^c \Gamma_7 \Phi + {\rm h.c.}
\label{eq:SA}
\eea
where the conjugate field is given by $\Psi^c = C \Gamma^0 \Psi^*$.
As fermion fields anticommute, it is also straightforward to show that the antisymmetric mass term
identically vanishes for a single field $\Psi = \Phi$ .

It is instructive to develop the Majorana mass terms Eqs.~(\ref{eq:SA}) using four dimensional
chiral components of $\Phi$ and $\Psi$.
We can label components of a six dimensional Dirac field according to their sign under
both $\Gamma_7$ and $\tilde{\Gamma}_5 = i \Gamma_0 \dots \Gamma_3$,
left- and right-handed chirality in 4D given by the projectors $P_{R,L} = (1 \mp \tilde{\Gamma}_5)/2$,
\be
\Psi = \left(
\ba{c}
\psi_{+R} \\ \psi_{+L} \\ \psi_{-L} \\ \psi_{-R}
\ea
\right) \quad .
\ee
So a 6D Dirac spinor is equivalent to two right-handed and two left-handed 4D Weyl spinors.
The Majorana mass terms in Eqs.~(\ref{eq:SA}) now become
\bea
&(S)& \quad \psi_{-R} \phi_{+R} + \psi_{+R} \phi_{-R}
-\psi_{-L} \phi_{+L} - \psi_{+L} \phi_{-L} \quad + {\rm h.c.} \nonumber \\
&(A)&	 \quad \psi_{-R} \phi_{+R} - \psi_{+R} \phi_{-R}
-\psi_{-L} \phi_{+L} + \psi_{+L} \phi_{-L} \quad + {\rm h.c.} \quad ,
\label{eq:SA2}
\eea
where we use the contracted product notation
$\psi_R \phi_R \equiv \psi_R^t (i\sigma_2) \phi_R = \epsilon^{ij} \psi_{R i} \phi_{R j}$
for right-handed spinors and
$\psi_L \phi_L \equiv \psi_L^t (-i\sigma_2) \phi_L = -\epsilon^{ij} \psi_{L i} \phi_{L j}$
for left-handed spinors, where $\epsilon^{ij}$ is the totally antisymmetric tensor of rank 2.
As $\psi \phi = \phi \psi$ (we recall that fermions anticommute),
it is now obvious that the antisymmetric mass term identically vanishes when $\Psi=\Phi$.
We see that a Majorana mass term in 6D always mixes ``+'' and ``$-$''
components. However, this does not {\it a priori} prohibit from building a
Majorana mass term for the vortex-bound left-handed neutrinos via a
see-saw mechanism, as the $+L$ components of the chiral zero modes
explicitly depend on the $-L$ components, therefore there are only two
degrees of freedom per zero mode~\cite{Libanov:2000uf}.

\subsection{Compactification on a sphere}

So let us consider a single fermion field $N$, singlet under the Standard Model gauge group, that will
play the role of the heavy ``right-handed" neutrino for the see-saw mechanism.
As discussed in the introduction, it is necessary to set the model with a compactification on a sphere
rather than with flat extra dimensions, in order to realize the see-saw mechanism in a consistent and realistic way.
The Lagrangian for a bulk field $N$ in $M^4 \times S^2$ with a (antisymmetric) Majorana mass term is given by
\be
\frac{\call_N}{\sqrt{-\det g_{AB}}}
= i \bar{N} \partial_\mu \Gamma^\mu N + \bar{N} \frac{\hat{D}}{R} N
-\frac{M}{2} (\bar{N}^c N + \bar{N} N^c) \quad ,
\label{eq:LN}
\ee
where the metric $g_{AB}$ is given by
\be
ds^2 = g_{AB} dx^A dx^B = g_{\mu \nu} dx^\mu dx^\nu - R^2 (d\theta^2 +
\sin^2 \theta d\phi^2) \quad ,
\ee
and
\be
\hat{D} = -i\Gamma ^{4}\left(\partial _{\theta }+\frac{\cot\theta }{2}
\right)-i\frac{\Gamma ^{5}}{\sin \theta }\partial _{\phi } \quad .
\ee
To obtain the 4D effective Lagrangian, the bulk field is decomposed into
orthonormal spinor spherical modes $\Upsilon_{lm}^\pm (\theta,\phi)$ (see
Ref.~\cite{Abrikosov:2002jr}) of the Dirac operator on a sphere
\be
N = \frac{1}{R} \sum_{\lambda,m} A \, \calu_{\lambda,m} \quad ,
\ee
where
\be A = \left( \ba{cc} \sigma^- \otimes 1 & \sigma^+ \otimes 1 \\
\sigma^+ \otimes 1 & \sigma^- \otimes 1 \\
\ea \right) \quad ,
\ee
with $\sigma^\pm = (\sigma_1 \pm i \sigma_2)/2$, and
\be
\hat{D}\cdot(A \, \calu_{\lambda,m})=i\lambda(A \, \calu_{\lambda,m})\quad,
\ee
with $\lambda = \pm (l+\frac{1}{2})$, $l =
\frac{1}{2}, \frac{3}{2},\dots$, $m = \pm \frac{1}{2}, \pm
\frac{3}{2},\dots$, $|m| \leq l$,
\be \calu_{\lambda,m}(x^\mu,\theta,\phi) = \left( \ba{c}
\Upsilon^\epsilon_{lm}(\theta,\phi) \otimes \xi_{\lambda,m}(x^\mu) \;
\e^{-i \pi/4} \\
\Upsilon^{-\epsilon}_{lm}(\theta,\phi) \otimes
\bar{\chi}_{\lambda,m}(x^\mu) \; \e^{i \pi/4} \\
\ea \right) \quad , \ee
where $\xi$ and $\bar{\chi} \equiv i\sigma_2 \chi^{*}$ are respectively
left-handed and right-handed Weyl spinors in Minkowski space (so both
$\xi$ and $\chi$ are left-handed). With this decomposition, we obtain the
4D effective Lagrangian
\bea \call_{eff} &=& \int d\theta \, d\phi \; \call_N = \sum_{\lambda,m}
\chi_{\lambda,m} i \partial \bar{\chi}_{\lambda,m} + \bar{\xi}_{\lambda,m}
i \bar{\partial} \xi_{\lambda,m} \nonumber \\
&& - \frac{\lambda}{R} \chi_{\lambda,m} \xi_{-\lambda,m} +
\frac{\tilde{M}}{2} (\xi_{\lambda,m} \xi_{-\lambda,-m} - \chi_{\lambda,m}
\chi_{-\lambda,-m} ) + {\rm h.c.} \quad ,
\label{eq:leff}
\eea
where $\partial \equiv \partial_\mu \sigma^\mu$, $\bar{\partial} \equiv
\partial_\mu \bar{\sigma}^\mu$, and $\tilde{M} = \epsilon(\lambda)
(-1)^{l-m} M$, $\epsilon (\lambda )\equiv \mbox{sign}(\lambda )$.

The equations of motion show that modes are related by groups of four when $M \neq 0$,
\be
\left(
\ba{cccc}
-\tilde{M} & 0 & -\lambda/R & k \\
0 & -\tilde{M} & \bar{k} & -\lambda/R \\
\lambda/R & k & -\tilde{M} & 0 \\
\bar{k} & \lambda/R & 0 & -\tilde{M} \\
\ea
\right)
\left(
\ba{c}
\chi_{-\lambda,-m} \\
\bar{\xi}_{\lambda,-m} \\
\xi_{-\lambda,m} \\
\bar{\chi}_{\lambda,m} \\
\ea
\right)
 = 0 \quad ,
\label{eq:motion}
\ee
with $i \partial = k$, $i \bar{\partial} = \bar{k}$.
We therefore obtain the correct dispersion relation
\be
k^2 = M^2 + \frac{\lambda^2}{R^2}
\ee
Notice that because of the positive curvature of the sphere, there are no zero-modes.
The propagator in momentum space is obtained by inverting the matrix operator in Eq.~(\ref{eq:motion})
\bea
\left< 0 \left| T \left(
\ba{cccc}
\chi_{--} \chi_{++} & \chi_{--} \bar{\xi}_{-+} & \chi_{--} \xi_{+-} & \chi_{--} \bar{\chi}_{--} \\
\bar{\xi}_{+-} \chi_{++} & \bar{\xi}_{+-} \bar{\xi}_{-+} & \bar{\xi}_{+-} \xi_{+-} & \bar{\xi}_{+-} \bar{\chi}_{--} \\
\xi_{-+} \chi_{++} & \xi_{-+} \bar{\xi}_{-+} & \xi_{-+} \xi_{+-} & \xi_{-+} \bar{\chi}_{--} \\
\bar{\chi}_{++} \chi_{++} & \bar{\chi}_{++} \bar{\xi}_{-+} & \bar{\chi}_{++} \xi_{+-} & \bar{\chi}_{++} \bar{\chi}_{--} \\
\ea
\right)
\right| 0 \right> = \nonumber \\
= \frac{i}{k^2-\tilde{M}^2-\lambda^2/R^2} \left(
\ba{cccc}
\tilde{M} & 0 & -\lambda/R & k \\
0 & \tilde{M} & \bar{k} & -\lambda/R \\
\lambda/R & k & \tilde{M} & 0 \\
\bar{k} & \lambda/R & 0 & \tilde{M} \\
\ea
\right) \quad ,
\label{eq:prop}
\eea
where abbreviated indices $\pm\pm$ stand for $\pm \lambda, \pm m$.

So to summarize, the six dimensional bulk field $N$ with a Majorana mass term is equivalent to a tower
of massive modes $\xi_{\lambda,m}$ and $\chi_{\lambda,m}$ in the 4D point of view,
with masses and propagators given by Eq.~(\ref{eq:prop}).

\section{See-saw mechanism with chiral zero modes}
\label{sec:see-saw}

\begin{table}
\bc
\bt{|rc|c|c|c|c|c|}
\hline
\multicolumn{2}{|c|}{fields}
& profiles & \multicolumn{2}{|c|}{charges} &
\multicolumn{2}{|c|}{representations} \\
\cline{4-7}
&&& $U(1)_g$ & $U(1)_Y$ & $SU(2)_W$ & $SU(3)_C$ \\
\hline
scalar & $\Phi$ & $F(\theta) \e^{i\phi}$ & +1 & 0 & {\bf 1} & {\bf 1} \\
&& $F(0)=0$, $F(\pi)=C_{\pi F}$ &&&& \\
\hline
scalar & $X$ & $X(\theta)$ & +1 & 0 & {\bf 1} & {\bf 1} \\
&& $X(0)=v_X$, $X(\pi)=0$ &&&& \\
\hline
scalar & $H$ & $H(\theta)$ & $-1$ & $+1/2$ & {\bf 2} & {\bf 1} \\
&& $H(0)=v_H$, $H(\pi)=0$ &&&& \\
\hline
fermion & $L_+$, $L_-$ & 3 L zero modes & $(3,0)$ & $-1/2$ & {\bf 2} & {\bf 1} \\
\hline
fermion & $E_+$, $E_-$ & 3 R zero modes & $(0,3)$ & $-1$ & {\bf 1} & {\bf 1} \\
\hline
fermion & $N$ & massive modes $\chi_{\lambda,m}$, $\xi_{\lambda,m}$ & 0 & 0 & {\bf 1} & {\bf 1} \\
\hline
scalar & $S_+$ & (composite field) & -1 & 0 & {\bf 1} & {\bf 1} \\
\hline
scalar & $S_-$ & (composite field) & 2 & 0 & {\bf 1} & {\bf 1} \\
\hline
\et
\ec
\caption{Field content of the model (scalars and leptons only).}
\label{table:fields}
\end{table}
The goal of this section is to implement the see-saw mechanism in a minimal model
with a vortex background on $M^4 \times S^2$.
Before venturing into this model with a full account of its technicalities, let us give
a brief argument on why the see-saw mechanism in 6D might explain the presence of large mixing angles
in the leptonic mixing matrix $U_{MNS}$.
As shown in Sec.~\ref{sec:majorana} (Eq.~(\ref{eq:SA2})), a Majorana mass term in 6D always connects
a ``+'' component with a ``$-$'' one.
A see-saw mechanism in 6D would amount to create an effective Majorana mass term for the lepton doublet $L$
(which contains the Standard Model neutrinos in the dimensional reduced effective theory) in a term of the form
\be
\bar{L^c} (A+B \, \Gamma_7) L \quad + {\rm h.c.} \quad ,
\label{eq:LcL}
\ee
for some coefficient $A,B$. The field $L$ is bound to the vortex, and gives rise to three left-handed
zero-modes $L_n$ ($n=1, 2, 3$ is therefore the family (or generation) number in 4D), 
with the $+L$ components explicitly dependent on the $-L$ ones.
Dropping the angular factor around the vortex (which does not change the argument), we have
\be
L_n \sim \left(
\ba{c}
0\\ f_2(n) \; l_n \\ f_3(n) \; l_n \\ 0
\ea
\right) \quad ,
\label{eq:Ln}
\ee
where $f_2$ and $f_3$ are functions of the extra dimensions that have a leading
behavior $f_2(n) \sim \theta ^{3-n}$ and $ f_3(n) \sim \theta ^{n-1}$
at a small distance $\theta$ near the core of the vortex.
As a result, the neutrino see-saw matrix element $(n,m)$ is expected to behave as
$f_2(n) f_3(m) + f_3(n) f_2(m) \sim \theta ^{2-|n-m|}$ (the see-saw matrix is
symmetric). Therefore, it will have a dominant entry at position $(1,3)$
and $(3,1)$
\be
M_\nu \sim \left(
\ba{ccc}
\cdot & \cdot & \times \\
\cdot & \cdot & \cdot \\
\times & \cdot & \cdot
\ea
\right) \quad ,
\ee
which guarantees one large, close to maximal mixing angle.
We also expect light neutrino masses to follow a ``pseudo-Dirac" inverted hierarchy mass pattern
$m_1 \simeq -m_2 \gg m_3$.
To have two large mixing angles in $U_{MNS}$ in a pattern close to the observed one, 
there are two possibilities.
Either one large mixing angle is attributed to the charged lepton mass matrix (more precisely in the
$2-3$ block in case it corresponds to the heaviest charged lepton states), and this
possibility is analyzed in Sec.~\ref{sec:ex1}, or both large mixing angles stem from the neutrino sector.
We will see in Sec.~\ref{sec:ex2} how this second possibility can emerge without contradicting
the simple line of reasoning outlined above. 
We now go back to the chosen setup on a sphere, and see how the model is implemented in details.

\subsection{Field content of the model}

The field content is similar to the model of Ref.~\cite{Frere:2001ug}, and
is given in Table~1 for easy reference
(we only consider the leptonic sector).
The brane is the Abelian vortex made of a gauge field $A$ (for the gauge group $U(1)_g$)
and a scalar field $\Phi = F(\theta) \e^{i\phi}$.
The scalar field $H$ has the quantum numbers of the Standard Model scalar doublet,
while $X$ is an additional scalar field needed to have inter generation mixings among quarks and leptons.
With a suitable scalar potential $V(\Phi,H,X)$ (see
Ref.~\cite{Libanov:2002tm}
where the flat space analogous case is discussed), the interaction with the vortex results in their localization
around the brane. The electroweak symmetry is spontaneously broken in the usual way by the field $H$,
while the vortex structure is generated by $\Phi$.

The charges of the fermions under the ``vortex" group $U(1)_g$ are now $(3,0)$ for
$(L_+,L_-)$ and $(0,3)$ for $(E_+,E_-)$. This differs from the ``flat case",
where we had taken half integer chiral charges, like +3/2 for $L_+$ and
$-3/2$ for $L_-$. The reason for the change is that on a sphere,
half-integer axial charges for fermions are inconsistent with the Dirac's
charge quantization condition~\cite{Frere:2003yv}. The interaction of
these fermions with the vortex field,
\be
g_l \Phi^3\bar L \frac{1-\Gamma_7}{2} L +
g_e \Phi^{*3}\bar E \frac{1-\Gamma_7}{2} E + {\rm h.c.} \quad ,
\label{eq:loc}
\ee
results in the localization of three chiral zero modes.
The bulk fermion singlet $N$ is not given any chiral charge under $U(1)_g$ so
that no term like Eq.~(\ref{eq:loc}) which would result in the localization of the field can be written.
In this paper, for simplicity, we choose $N$ to have no charge under $U(1)_g$.
Otherwise, the decomposition of the bulk field into spherical modes given in Sec.~\ref{sec:majorana}
would not be valid.

In Table~1, $S_+$ and $S_-$ are not necessarily additional elementary scalar fields in the model,
but represent effective couplings lumping products of the existing scalars
already introduced, that leave the following Lagrangian invariant
($\tilde{H}=i\sigma _{2}H^*$)
\be
\frac{\call_D}{\sqrt{-\det g_{AB}}} = \sum \limits_{S_{+} }^{}Y^+_\nu(S_+)
\tilde{H} S_+ \bar{L} \frac{1 + \Gamma_7}{2} N + \sum
\limits_{S_{-}}^{}Y^-_\nu (S_-) \tilde{H} S_- \bar{L} \frac{1 - \Gamma_7}{2} N + {\rm h.c.} \quad ,
\label{eq:LD}
\ee
therefore $S_+$ and $S_-$ can be
\bea
S_+ &=& \Phi^*, \; X^*, X^{*2} \Phi, \; \dots \nonumber \\
S_- &=& X^2, \; X \Phi, \; \Phi^2, \; \dots
\label{eq:Spm}
\eea
Notice that we don't limit the Lagrangian to quartic terms, as
the 6D theory is not renormalizable, and can only be seen as the low-energy part
of a more complex structure, so $Y^\pm_\nu$ are dimensionful couplings, as will be later discussed.
The reduction of the Lagrangian $\call_D$ to four dimensions will give rise to Dirac type masses for the
Standard Model neutrinos.
A non-vanishing result appears only for terms with a total winding number around the vortex equal to zero.
This leads to selection rules among the modes of the bulk field $N$.
Therefore, a different flavor structure of the see-saw mass matrix for the three
light neutrinos arises for each $S_+$ and $S_-$, as they have different winding numbers around the vortex.

\subsection{Chiral zero-modes and neutrino Dirac masses}

Fermionic zero-modes in a vortex background on a sphere have been
calculated in Ref.~\cite{Frere:2003yv}.
Using the same notations, the three zero for $L$ are written as
\be
L_n(\theta,\phi,x^\mu) =
\left(
\ba{c}
0\\
\e^{-i \phi (n-7/2)} f_2(n,\theta) \; l_n(x^\mu) \\
\e^{-i \phi (n-1/2)} f_3(n,\theta) \; l_n(x^\mu) \\
0
\ea
\right) \quad ,
\ee
with $n = 1,2,3$, and $l_n$ is a two component spinor.
Let $s_\pm$ be the winding number of the composite scalar fields $S_\pm$, so
\be
S_\pm(\theta,\phi) = S_\pm(\theta) \; \e^{i s_\pm \phi} \quad .
\ee
Following Sec.~\ref{sec:majorana}, the bulk field $N$ is decomposed in a tower of spherical modes.
In a more explicit fashion, we have
$\Upsilon^\epsilon_{lm}(\theta,\phi) = \frac{\e^{i m \phi}}{\sqrt{2\pi}}
\left(
\ba{c}
S^\epsilon_{u,lm}(\theta) \\
S^\epsilon_{d,lm}(\theta)
\ea
\right)$, so
\be
N(\theta,\phi,x^\mu) = \sum_{\lambda,m} \frac{\e^{i m \phi}}{\sqrt{2\pi}R} \left(
\ba{c}
S^{-\epsilon}_{d,lm}(\theta) \; \e^{i\pi/4} \; \bar{\chi}_{\lambda,m}(x^\mu) \\
S^\epsilon_{u,lm}(\theta) \; \e^{-i\pi/4} \; \xi_{\lambda,m}(x^\mu) \\
S^\epsilon_{d,lm}(\theta) \; \e^{-i\pi/4} \; \xi_{\lambda,m}(x^\mu) \\
S^{-\epsilon}_{u,lm}(\theta) \; \e^{i\pi/4} \; \bar{\chi}_{\lambda,m}(x^\mu)
\ea
\right) \quad .
\ee
By integrating over the extra dimensions, we get the effective neutrino Dirac mass Lagrangian
$\int d\theta \, d\phi \call_D \equiv \call_+ + \call_-$,
\be
\call_\pm = \sum_{n,s_\pm,\lambda} M_D^\pm(\lambda,n,s_\pm) \; \bar{l}_n \bar{\chi}_{\lambda,m_\pm}
+ {\rm h.c.} \quad ,
\label{eq:ldir}
\ee
with the selection rules from the $\phi$ integration
\bea
m_+ &=& \frac{1}{2} -n -s_+ \nonumber \\
m_- &=& \frac{7}{2} -n -s_- \quad ,
\eea
and
\bea
M_D^+(\lambda,n,s_+) &=& \int d\theta \, \sin \theta Y_\nu^+(S_+) H(\theta) S_+(\theta)(\sqrt{2\pi} R f_3(n,\theta))
S^{-\epsilon}_{d,l,m_+}(\theta) \, \e^{i\pi/4} \nonumber \\
M_D^-(\lambda,n,s_-) &=& \int d\theta \, \sin \theta Y_\nu^-(S_-) H(\theta) S_-(\theta)(\sqrt{2\pi} R f_2(n,\theta))
S^{-\epsilon}_{u,l,m_-}(\theta) \, \e^{i\pi/4} \quad .
\label{eq:ynu}
\eea

\subsection{Neutrino see-saw masses when $M = 0$}
\label{sec:M0}

When $M=0$, lepton number is a conserved quantity in the Lagrangians Eqs.~(\ref{eq:leff}) and~(\ref{eq:ldir}).
Therefore, as $M_D^\pm \ll 1/R$, the three light neutrino mass eigenstates are exactly massless,
and mixings of flavor eigenstates with massive states are suppressed.

In order to generate light neutrino masses, one might introduce explicit lepton number violating mass terms.
In analogy with Eq.~(\ref{eq:LD}), we can consider the following Lagrangian
\be
\frac{\call_{LV}}{\sqrt{-\det g_{AB}}} = \sum \limits_{S'_+}^{}Y'^+_\nu
\tilde{H} S'_+ \bar{L} \frac{1 + \Gamma_7}{2} N^c + \sum \limits_{S'_-}^{}
Y'^-_\nu \tilde{H} S'_- \bar{L} \frac{1 - \Gamma_7}{2} N^c + {\rm h.c.}
\quad ,
\label{eq:LLV}
\ee
where $S'_\pm$ are composite scalar fields with a winding number $s'_\pm$,
\bea
S'_+ &=& \Phi^*, \; X^*, X^{*2} \Phi, \; \dots \nonumber \\
S'_- &=& X^2, \; X \Phi, \; \Phi^2, \; \dots
\eea
It turns out that this attempt still fails to generate neutrino Majorana masses for the light states.
It can be checked that all three Lagrangians Eqs.~(\ref{eq:LN}),~(\ref{eq:LD}) and~(\ref{eq:LLV})
(with $M=0$) as well as the kinetic terms for $L$ are invariant under the discrete symmetry
\bea
N &\rightarrow& \e^{i \frac{\pi }{2} (\Sigma + \tilde{\Gamma}_5)} N
\nonumber \\
L &\rightarrow& \e^{i \frac{\pi }{2} (\Sigma - \tilde{\Gamma}_5)} L \quad ,
\label{eq:sym}
\eea
where $\tilde{\Gamma}_5 = i \Gamma_0 \dots \Gamma_3$, $\Sigma = i \Gamma_4 \Gamma_5$,
and a simultaneous sign flip of the coupling constant $g_{l}\to -g_{l}$.
The latter does not affect the zero-modes of $L$ and thereby is irrelevant in the following
discussion. The action of this symmetry is more readable when $N$ and $L$
are decomposed into their chiral components. Simply,
\bea
\psi \rightarrow -\psi \quad &{\rm if}& \quad \psi= N_{+R},\,
N_{+L},  \, L_{-R}, \, L_{-L} \nonumber \\
\psi \rightarrow \psi \quad &{\rm if}& \quad \psi= N_{-R}, \, N_{-L}, \, L_{+R}, \, L_{+L}
\eea
So when $M=0$, the Lagrangians considered so far are invariant under this symmetry.
However, a Majorana mass term for the light neutrinos corresponds to the effective coupling Eq.~(\ref{eq:LcL}),
which is not invariant under the transformation Eq.~(\ref{eq:sym}).

\subsection{Neutrino see-saw masses when $M \neq 0$}

A non-zero bulk Majorana mass term for $N$ does break the symmetry Eq.~(\ref{eq:sym}),
so that neutrino see-saw masses can indeed be generated in this case.
They are calculated as truncated two-point functions with a transfer momentum $k \rightarrow 0$.
For the see-saw mechanism to work, the ``magnetic" quantum numbers of the bulk
field modes $m_+$ and $m_-$ in Eq.~(\ref{eq:ynu})  have to be opposite in value. 
We obtain the following neutrino see-saw mass matrix
\be
M_\nu(n,m) = \langle \bar{l}_n \bar{l}_m\rangle  = \sum_{\lambda,s_+}
\frac{-M_D^+(\lambda,n,s_+) M_D^-(-\lambda,m,s_-)M(-1)^{l-(1/2-n-s_+)} \epsilon(\lambda)}{M^2 + \lambda^2/R^2}
+ n \leftrightarrow m \quad ,
\label{eq:mnuM}
\ee
where $n,m = 1,2,3$ are the generation indices, and a non-zero contribution appears only 
if the selection rule $n + m + s_+ + s_- = 4$ is respected.

\subsection{Charged lepton masses}

In the same spirit of the neutrino sector, we write down all possible interactions that give a mass to the
charged leptons, and investigate all possibilities
\bea
\frac{\call_E}{\sqrt{-\det g_{AB}}} =\sum \limits_{S_+^l}^{} Y^+_l(S_+^l)
S^l_+ H \bar{L} \frac{1 + \Gamma_7}{2} E + \sum
\limits_{S_-^l}^{}Y^-_l(S_-^l) S^l_- H \bar{L} \frac{1 - \Gamma_7}{2} E +
{\rm h.c.} \quad ,
\label{eq:LE}
\eea
where $S^l_\pm$ can be
\be
S^l_\pm = \dots, \; X^2 \Phi^*, \; X, \; \Phi, \; X^* \Phi^2, \; \dots
\ee
The interaction of $E$ with the vortex background Eq.~(\ref{eq:loc}) leads to the localization of three
right-handed zero-modes~\cite{Frere:2003yv}
\be
E_m(\theta,\phi,x^\mu) =
\left(
\ba{c}
\e^{-i \phi (m-1/2)} f_3(m,\theta) \; \bar{e}_m(x^\mu) \\
0\\
0\\
\e^{-i \phi (m-7/2)} f_2(m,\theta) \; \bar{e}_m(x^\mu)
\ea
\right) \quad , \quad m=1,2,3 \quad .
\ee
After integration over the extra dimensions, we get the charged lepton mass Lagrangian
$\int d\theta \, d\phi \call_E \equiv \call_+^l + \call_-^l$,
\be
\call^l_\pm = \sum_{n,m} M_l^\pm(n,m) \; \bar{l}_n \bar{e}_m + {\rm h.c.} \quad ,
\ee
with
\bea
M_l^+(n,m) &=& 2\pi R^2 \int d\theta \, \sin \theta Y_l^+(s_+) H(\theta) S_+(\theta)
f_3(n,\theta) f_3(m,\theta) \delta_{n-m+s_+,0} \nonumber \\
M_l^-(n,m) &=& 2\pi R^2 \int d\theta \, \sin \theta Y_l^-(s_-) H(\theta) S_-(\theta)
f_2(n,\theta) f_2(m,\theta) \delta_{n-m+s_-,0}
\label{eq:yl}
\eea
%

\section{Estimates for mass and mixing matrices}
\label{sec:matrices}

The Brout-Englert-Higgs field is localized on a narrow region of typical size
$R \theta_\Phi$, locked to the profile of $\Phi$ through the scalar
potential~\cite{Libanov:2005mv, Libanov:2007zz}. Therefore, integrals over
$\theta$ that appear in Eqs.~(\ref{eq:ynu}) and Eqs.~(\ref{eq:yl}) are
typically saturated at $\theta = \theta_\Phi \ll 1$. To estimate the mass
matrix elements, we therefore use approximate profiles for the various
fields that are accurate enough under this narrow BEH scalar assumption. On
$[0,\theta_\Phi]$, we have
\bea
H(\theta) &\simeq& H(0) = v_H \nonumber \\
X(\theta) &\simeq& X(0) = v_X \nonumber \\
F(\theta) &\simeq& \frac{\theta}{\theta_\Phi} C_{\pi F}
\label{eq:scalprof}
\eea
For the fermion zero modes, it has been shown in Ref.~\cite{Frere:2003yv}
that
on $[0,\theta_\Phi]$, the profiles  are approximately given by ($n=1,2,3$)
\bea
\sqrt{2\pi} R f_2(n, \theta) &\simeq& \frac{1}{\theta_A}
\left( \frac{\theta}{\theta_A} \right)^{3-n}
\left[ 1 + \frac{\theta_A}{\theta_\psi} \left( \frac{\theta_A}{\theta_\Phi} \right) \right]^{\delta_{n1}}
\nonumber \\
\sqrt{2\pi} R f_3(n, \theta) &\simeq& \frac{1}{\theta_\psi}
\left( \frac{\theta}{\theta_A} \right)^{n-1}
\left( \frac{\theta_A}{\theta_\Phi} \right)^{\delta_{n3}} \quad ,
\label{eq:fermprof}
\eea
with typically $\theta_\Phi \ll \theta_A \leq \theta_\psi$.

Operators with different dimensions in Eqs.~(\ref{eq:LD}), (\ref{eq:LLV}) and (\ref{eq:LE}) are unsuppressed
if $v_X \sim C_{\pi F} \sim \Lambda^2$, where $\Lambda$ is the energy scale of the model.
The value $v_H$ of the field $H$ in the core of the vortex cannot however be as large.
Indeed, as $H$ is charged under $SU(2)_W$, the relation
\be
2\pi R^2 \int_0^\pi d\theta \, \sin \theta H^2(\theta) = \frac{V_{SM}^2}{2} \quad
\label{eq:VSM}
\ee
which defines the effective Brout-Englert-Higgs scalar expectation value $V_{SM} \simeq 250$~GeV,
has to be satisfied~\cite{Libanov:2005mv, Libanov:2007zz}.

What is the ``new" scale $\Lambda$ of the model?
While the obvious dimensional parameter is $1/R$, the energy scale really appearing is determined
by the size of the vortex, rather than the sphere, namely $\Lambda \sim 1/(\theta_\Phi R)$, and $\theta_\Phi$
depends mostly on the fermion spectrum choices we
make~\cite{Libanov:2002tm, Libanov:2002ka, Libanov:2005mv, Libanov:2007zz}.
From pure phenomenological considerations, the model is constrained by
flavour violating processes, with the strongest constraint arising from
the non observation of the decay $K \rightarrow \mu^\pm e^\mp$; it
requires the size $R$ of the extra dimensions to satisfy $1/R \geq
64$~TeV~\cite{Frere:2003ye}. Just to fix ideas, we will assume here that
$1/R \sim 100$~TeV. In what follows, we take $\theta_\Phi \sim 0.1$, a
value that enables to reproduce quark and charged lepton hierarchical mass
patterns, so that $\Lambda \sim 10^3$~TeV, and $v_H \sim 10^2~{\rm
TeV}^2$.

\subsection{Charged lepton mass matrix}

If we assume that all terms in the Lagrangian $\call_E$ (Eq.~(\ref{eq:LE})) have order one
dimensionless coefficients, \ie if a generic term with winding number $s$ has a coupling
$Y_l^\pm = \tilde{Y}_l^\pm(s) \cdot \Lambda^{-(2|s|+2|s-1|+1)}$ with $\tilde{Y}_l^\pm(s) \sim {\cal O}(1)$,
then the charged lepton mass matrices $M_l^+$ and $M_l^-$ corresponding to terms with the projector
$(1+\Gamma_7)$ and $(1-\Gamma_7)$ resp. have the following structure
\be
M_l^+ \sim \frac{v_H}{\Lambda} \delta^2 \delta_A^2 \left(
\ba{ccc}
1 & \delta & \delta \\
\delta & \delta^2 & \delta^2 \\
\delta & \delta^2 & \delta^2
\ea
\right) \quad , \quad
M_l^- \sim \frac{v_H}{\Lambda} \delta^2 \left(
\ba{ccc}
\delta^4 \beta^2 & \delta^3 \beta & \delta^2 \beta \\
\delta^3 \beta & \delta^2 & \delta \\
\delta^2 \beta & \delta & 1
\ea
\right) \quad ,
\label{eq:ml}
\ee
where $\delta = \theta_\Phi/\theta_A$, $\delta_A=\theta_A/\theta_\psi$ and $\beta=(1+\delta_A/\delta)$.
Notice that these matrices are in general not symmetric.

The matrices in Eq.~(\ref{eq:ml}) exhibit a hierarchical structure which
can accommodate the observed hierarchy of the charged lepton masses $m_e
\ll m_\mu \ll m_\tau$. Large mixing angles can also be present. Indeed,
$M_l^+$ can easily harbor a large mixing angle in the $2-3$ block, and if
$\beta \simeq 1/\delta$, $M_l^-$ can also have one in the $1-2$ block.
However, such a large mixing between the lightest charged lepton mass
eigenstates cannot lead to the observed $U_{MNS}$ matrix. Instead, the
large mixing has to occur among the heaviest mass eigenstates. If only
$M_l^-$ ($M_l^+$) is present, this requires a tuning of the coefficients
in the ratio $\tilde{Y}_l^-(0)/\tilde{Y}_l^-(1) \simeq \delta/2$ (resp.
$\tilde{Y}_l^+(0)/\tilde{Y}_l^+(-1) \simeq \delta/2$). It is then possible
to accommodate both the observed charged lepton masses and a large mixing
angle between $\mu$ and $\tau$. The presence of a factor $\beta$ in
$M_l^-$ enables to achieve hierarchical masses more easily compared to
$M_l^+$. As a result, our numerical study showed that for $M_l^+$, it is
not possible to have one large and two small mixing angles in the mixing
matrix $U_l$. For $M_l^-$, on the other hand, with $\theta_\Phi \sim 0.1$,
$\theta_A \simeq 1$, $\theta_\psi = \pi$, we could obtain the observed
charged lepton mass ratios $m_\tau/m_\mu$, $m_\mu/m_e$, together with a
maximal mixing between $\tau$ and $\mu$; the small mixing angles in $U_l$
are then of order $\delta$. Also, $m_\tau$ has naturally the right order
of magnitude in our model if $\tilde{Y}_l^- \sim 1$ as $m_\tau \sim
\delta^2 v_H / \Lambda \sim 1$~GeV. When both $M_l^+$ and $M_l^-$ are
present, an even stronger fine-tuning of the coefficients is needed to
achieve a large mixing angle, so we discard this case. In summary, one can
accommodate the observed hierarchy of the charged lepton masses  $m_e \ll
m_\mu \ll m_\tau$ and one large mixing angle (that observed in atmospheric
neutrinos oscillations) in this model, when only operators with the
projector $(1-\Gamma_7)$ are present in the Lagrangian Eq.~(\ref{eq:LE}).

\subsection{Neutrino see-saw mass matrix when $M \neq 0$}
\label{sec:Mnot0}

To estimate the neutrino see-saw masses of Eq.~(\ref{eq:mnuM}), where the Dirac masses $M_D^\pm$
are given by Eqs.~(\ref{eq:ynu}),
we recall the explicit expression of the spherical functions $S_{u,lm}^\epsilon$ and $S_{d,lm}^\epsilon$
\bea
\left(
\ba{c}
S_{u,lm}^\epsilon (\theta) \\
S_{d,lm}^\epsilon (\theta)
\ea
\right) &=&
\epsilon \, i^{l^+} \, (-1)^{\frac{m^-+|m^-|}{2}} \frac{\sqrt{(l+m)!(l-m)!}}{2^{|m|+1/2}\Gamma(l^+)}
\times \nonumber \\
&\quad& \quad \times \,
\left(
\ba{c}
\e^{-i\pi \epsilon/4} \rho^{|m^-|,|m^+|}(x) P^{|m^-|,|m^+|}_{l-|m|}(x) \\
\epsilon(m) \e^{i\pi \epsilon/4} \rho^{|m^+|,|m^-|}(x) P^{|m^+|,|m^-|}_{l-|m|}(x)
\ea
\right) \quad ,
\label{eq:susd}
\eea
where $l^\pm = l \pm 1/2$, $m^\pm = m \pm 1/2$, $\epsilon
=\mbox{sign}(\lambda ) $, $x=\cos \theta$,
$\rho^{\alpha,\beta}(x)=\sqrt{(1-x)^\alpha (1+x)^\beta}$ and
$P^{\alpha,\beta}_n$ are the Jacobi polynomials, and notice the following
points:
\bi
\item All the terms in Eq.~(\ref{eq:mnuM}) have a phase equal to one. This
can be easily checked using expressions in Eq.~(\ref{eq:susd}).
\item For $0 \leq \theta \leq \theta_\Phi \ll 1$~\cite{Bateman:1953},
\bea
|S_{u,lm}^\epsilon| &\sim& (l \theta)^{|m^-|} \sqrt{l} \quad {\rm at} \quad l<1/\theta \nonumber \\
|S_{d,lm}^\epsilon| &\sim& (l \theta)^{|m^+|} \sqrt{l} \quad {\rm at} \quad l<1/\theta \nonumber \\
|S_{u,lm}^\epsilon|, \, |S_{d,lm}^\epsilon| &\sim& \frac{\cos(l\theta)}{\sqrt{\theta}}
\quad {\rm at} \quad l \geq 1/\theta
\eea
As a result, for $l \gg 1/\theta_\Phi$, the Dirac masses $M_D^\pm$ are increasingly suppressed.
Therefore, one can limit the sum in Eq.~(\ref{eq:mnuM}) to values of $\lambda$ that are smaller to
$\lambda_{max} \sim 2\pi/\theta_\Phi$ in module.
\ei
As for the charged leptons, we express the dimensionful Yukawa couplings $Y_\nu^\pm$ in terms of 
a dimensionless coefficient $\tilde{Y}_\nu^\pm$ and a power of the energy scale $\Lambda$.
From Eq.~(\ref{eq:Spm}), we have 
$Y_\nu^+(S_+) = \tilde{Y}_\nu^+(s_+) \cdot \Lambda^{-(2|s_+|+2|s_+ +1|+1)}$ and 
$Y_\nu^-(S_-) = \tilde{Y}_\nu^-(s_-) \cdot \Lambda^{-(2|s_-|+2|s_- -2|+1)}$. 
Assuming $v_X \sim C_{\pi F} \sim \Lambda^2$, 
we obtain the following structure for the neutrino (symmetric) see-saw mass matrix
\be
M_\nu \sim \frac{v_H^2}{\Lambda^2} \frac{M}{M^2+1/R^2} \delta^3 \delta_A \left(
\ba{ccc}
\beta \delta^2 & \delta & 1 \\
\delta & \delta^2 & \delta \\
1 & \delta & \delta^2
\ea
\right) \quad .
\label{eq:Mnu}
\ee
The light neutrinos masses follow the inverted hierarchy pattern $|m_1| \simeq |m_2| \gg |m_3|$,
with $|m_3| \sim \delta^2 |m_1|$.
Moreover, $m_1$ and $m_2$ form a ``pseudo-Dirac" pair as $m_1 + m_2 \sim \delta^2 |m_1|$.
Therefore, this model naturally predicts a hierarchy in the mass squared splittings relevant
in neutrino oscillation experiments $\Delta m_{21}^2/\Delta m_{13}^2 \sim \delta^2$,
in good agreement with the observed data
$\Delta m_{21}^2/\Delta m_{13}^2 \simeq
3.2\%$~\cite{GonzalezGarcia:2010er}.

For the neutrino masses to be in the sub-eV range, we need a Majorana mass that is either very large,
$M \geq 10^{11}$~GeV, or very small, $M \leq 10$~GeV, compared to the compactification scale.
A third possibility is to trade this large or small Majorana mass for smaller Yukawa couplings.
In this case, we can suppose that the Majorana mass $M \sim 1/R$,
which brings all dimensional quantities to their natural scale.
Also, it is more natural to suppose that $M \sim 1/R$ rather than $M \sim \Lambda$, as the field $N$
does not interact with the vortex.
A value around $10^{-3}$ for the dimensionless couplings in Eq.~(\ref{eq:LD}) gives
$\Delta m^2_{13} \simeq 2.5 \cdot 10^{-3}~{\rm eV}^2$.

The neutrino mass matrix Eq.~(\ref{eq:Mnu}) is diagonalized by a matrix $U_\nu$ with the structure
\be
U_\nu \sim \left(
\ba{ccc}
1/\sqrt{2} & 1/\sqrt{2} & \delta \\
\delta & \delta & 1 \\
-1/\sqrt{2} & 1/\sqrt{2} & \delta
\ea
\right) + {\cal O}(\delta^2) \quad .
\ee
Let us emphasize that the large mixing angle in the $1-3$ block is maximal up to $\delta^2$ corrections.
When the charged lepton mass matrix contains a large mixing angle in the $2-3$ block, this model
predicts two large mixing angles in $U_{MNS}=U_l^\dagger U_\nu$, as observed.
The remaining small mixing angle $U_{e3}$, which corresponds to the weight of the lightest mass eigenstate
in the electronic neutrino, is predicted to be of order $\delta$.
Moreover, as all phases in Eq.~(\ref{eq:mnuM}) are real, there is no $CP$ violation in this model.

As the observed $\theta_{13}$ angle in Eq.~(\ref{eq:umns}) is small,
the matrix $U_{MNS}$ is often parameterized in the following form (neglecting possible $CP$ phases)
\be
U_{MNS} \simeq \left(
\ba{ccc}
\cos \theta_\odot & \sin \theta_\odot & \epsilon \\
-\cos \theta_\oplus \sin \theta_\odot & \cos \theta_\oplus \cos \theta_\odot & \sin \theta_\oplus \\
\sin \theta_\oplus \sin \theta_\odot & -\sin \theta_\oplus \cos \theta_\odot & \cos \theta_\oplus
\ea
\right) \quad ,
\label{eq:UMNSapprox}
\ee
where the solar (neutrino) angle $\theta_\odot \simeq 35^\circ$,
the atmospheric (neutrino) angle is close to maximal $\theta_\oplus \simeq 45^\circ$ and
$|\epsilon| \ll 1$.
This form suggests that $\theta_\oplus$ is due to $U_l$ while $\theta_\odot$ originates from $U_\nu$.
In our model, it is amusing to notice that the almost maximal value of $\theta_\oplus$ is ``accidental",
whereas the non-maximal value of $\theta_\odot$ is due to a shift of the quasi-maximal angle
in $U_\nu$ by small mixing angles $\sim \delta$ present in $U_l$, when the product
$U_{MNS} = U_l^\dagger U_\nu$ is performed.

\subsection{Numerical example one}
\label{sec:ex1}

We found that a value $\delta = 0.07$ is suitable to accommodate both the steep hierarchy of charged
lepton masses, and the large (but not maximal for $\theta_{12}$) mixing angles in $U_{MNS}$.
So we took $\theta_\Phi = 0.07$, $\theta_A = 1$, $\theta_\psi = \pi$, and the energy scale is fixed at
$\Lambda = 10^3$~TeV. For the scalar fields, as discussed earlier, we considered that $v_X = C_{\pi F} = \Lambda^2$.
For $v_H$, using $V_{SM} = 250$~GeV, Eq.~(\ref{eq:VSM}) gives $v_H \simeq 100~{\rm TeV}^2$.

For the charged lepton mass matrix $M_l$, only interactions with projector $(1-\Gamma_7)$ in Eq.~(\ref{eq:LE})
are kept, with coefficients $\tilde{Y}_l^- = y_0 \, \{0.1,3.4,1,30,0.1\}$ for operators with a winding number
$s = -2, \dots , 2$ with the factor $y_0 = 5.0$ fixed by the $m_\tau$
mass\footnote{These large values of $\tilde{Y}_l^-$ are still within the 
perturbative range if factors of $(2\pi)$
are taken into account. For instance, the effective coupling for $s=1$ is
$\tilde{Y}_l^-(1)/(2\pi)^3 \simeq 0.6$. Also, $v_X$ and $C_{\pi F}$ could be raised above $\Lambda^2$ to decrease
the coefficients.}.
We obtain
\be
M_l = \left(
\ba{ccc}
3.01 \cdot 10^{-4} & 2.33 \cdot 10^{-2} & 1.11 \cdot 10^{-3} \\
2.64 \cdot 10^{-3} & 3.0 \cdot 10^{-3} & 1.28  \\
1.11 \cdot 10^{-3} & 1.46 \cdot 10^{-1} & 1.22
\ea
\right) \quad [{\rm GeV}] \quad .
\ee
This matrix does lead to the observed charged lepton masses, although this requires some tuning
in the coefficients $\tilde{Y}_l^-$. $M_l$ is diagonalized by two unitary matrices $U_l$ and $V_l$
\be
U_l^\dagger M_l V_l = D_l = {\rm diag}\{m_e, m_\mu, m_\tau \} \quad ,
\ee
with
\be
U_l = \left(
\ba{ccc}
0.976 & -0.219 & 0.0014 \\
0.151 & 0.676 & 0.722  \\
-0.159 & -0.704 & 0.692
\ea
\right) \quad , \quad
V_l = \left(
\ba{ccc}
0.999 & -0.0088 & -0.0015 \\
0.0087 & 0.998 & 0.058  \\
-0.002 & -0.058 & 0.998
\ea
\right) \quad .
\ee
So $U_l$ contains one large mixing angle in the $2-3$ block, tuned to be close to maximal,
while $V_l$ is close to identity.

\bfig[t]
\bc
\includegraphics[width=0.7\columnwidth, clip]{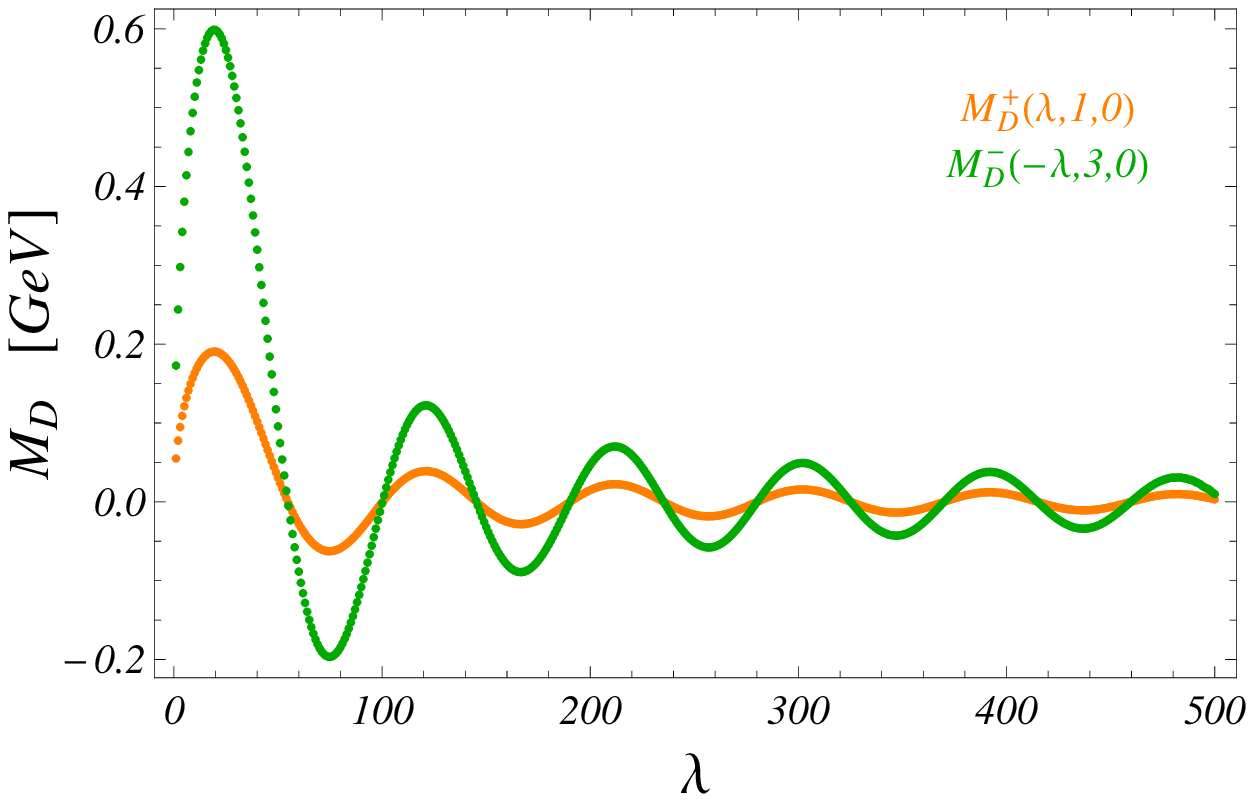}
\caption{Dirac masses given by Eq.~(\ref{eq:ynu}) as a function of $\lambda$,
with $\Lambda = 100$~TeV, $C_{\pi F} = v_X = \Lambda^2$,
$\theta_\Phi = 0.07$, $\theta_A = 1$, $\theta_\psi = \pi$,
and dimensionless coefficients $\tilde{Y}_l^-$ equal to 1.
}
\ec
\efig
For neutrinos, the Dirac masses calculated with Eq.~(\ref{eq:ynu}) decrease rather slowly with $\lambda$,
as can be seen in Fig.~1. Therefore, when $M$ is large compared to $1/R$,
a large number of modes is needed to calculate neutrino see-saw masses with some precision.
On the other hand, as discussed in Sec.~\ref{sec:Mnot0}, it is preferable to suppose a
Majorana mass around the compactification scale $M \sim 1/R$. We take $M=1/R=70$~TeV.
When $M \sim 1/R$, the number of modes that contribute significantly to
neutrinos masses is much more limited, because of the increasing curvature mass term in the denominator of
Eq.~(\ref{eq:mnuM}). In this numerical example, we take
$\lambda_{max}=100$. We have checked that taking a larger value of 
$\lambda_{max}$ ($=500$) doesn't affect the results presented here.
For all the operators with the lowest dimension in
Eq.~(\ref{eq:LD}), corresponding to winding number $s_+ = \{-1,0\}$ and
$s_- = \{0,1, 2\}$, we take a common value $y_\nu$ for the
dimensionless coefficient; all higher dimensional operators are neglected.
The value $y_\nu = 7.5 \cdot 10^{-3}$ is fixed by the observed $\Delta m^2_{31}$ for
atmospheric neutrinos. We find that these parameters give a good fit to all
neutrino data, with the correct value of $\Delta m^2_{21}$ for solar neutrinos following automatically. 
We obtain
\be
M_\nu = \left(
\ba{ccc}
7.78 \cdot 10^{-4} & 1.65 \cdot 10^{-3} & 4.74 \cdot 10^{-2} \\
1.65 \cdot 10^{-3} & 1.30 \cdot 10^{-4} & 1.53 \cdot 10^{-3} \\
4.74 \cdot 10^{-2} & 1.53 \cdot 10^{-3} & 0
\ea
\right) \quad {\rm [eV]} \quad .
\ee
$M_\nu$ is diagonalized by one unitary matrix $U_\nu$
\be
U_\nu^\dagger M_\nu U_\nu^* = D_\nu =
{\rm diag}\{-4.699 \cdot 10^{-2}, 4.787 \cdot 10^{-2}, 2.358 \cdot 10^{-4} \} \quad {\rm [eV]} \quad ,
\ee
with
\be
U_\nu = \left(
\ba{ccc}
0.704 & 0.709 & -0.032 \\
-0.0016 & 0.047 & 0.999 \\
-0.710 & 0.703 & -0.034
\ea
\right) \quad .
\ee
As announced in Sec.~\ref{sec:see-saw}, $U_\nu$ contains one large, close to maximal mixing angle,
and the neutrino mass spectrum is pseudo-Dirac.
We obtain $\Delta m^2_{13} = 2.21 \times 10^{-3}~{\rm eV^2}$, $\Delta m^2_{21} = 8.39 \times 10^{-5}~{\rm eV^2}$,
so $\Delta m^2_{21}/\Delta m^2_{13} = 3.8 \%$, all these values are close
to the observed ones.

Finally, for the matrix $U_{MNS} = U_l^\dagger U_\nu$, we get
\be
U_{MNS} = \left(
\ba{ccc}
0.800 & 0.587 & 0.124 \\
0.344 & -0.619 & 0.706 \\
0.492 & -0.522 & -0.697
\ea
\right) \quad .
\ee
Therefore $\tan^2 \theta_{12} = 0.539$, $\tan^2 \theta_{23} = 1.026$, $\sin^2 \theta_{13} = 1.55 \cdot 10^{-2}$.
The present matrix is of course real, as, for simplicity we did not try to
include CP violation; this can of course be considered as a future extension.

The amplitude for neutrinoless double-beta decay ($0\nu \beta \beta$) is proportional to a
quantity commonly referred to as the effective neutrino Majorana mass,
$\left< m_{\beta \beta} \right> = \sum_i m_i U_{e i}^2$. In our example, we have
\be
|\left< m_{\beta \beta} \right>| = 13.5 \quad {\rm meV} \quad ,
\ee
with a partial cancellation due to the pseudo-Dirac pattern.

Although the precise values obtained here should not be taken too seriously,
as they rely on rough approximations for the scalar and fermion fields in extra-dimensions,
nevertheless, this example shows how the puzzling pattern experimentally observed in
the $U_{MNS}$ matrix can arise in our model, while ensuring hierarchical charged lepton masses.
Moreover, the prediction of an interesting pseudo-Dirac structure for neutrinos may be of great
experimental significance.

\subsection{Numerical example two}
\label{sec:ex2}

When light neutrino masses have an inverted hierarchy with a pseudo-Dirac pair, 
a particular structure of the neutrino see-saw matrix can give rise naturally to the presence of 
two large mixing angles in the $U_{MNS}$ matrix~\cite{Datta:2003qg}.
Indeed, if $-m_1 \simeq m_2 \gg m_3$ and $U_{MNS}$ is given by Eq.~(\ref{eq:UMNSapprox}) with 
$\theta_\odot = \pi/4$, we are led to the following pattern at leading order
\be
M_\nu  = U_{MNS} \cdot D_\nu \cdot U_{MNS}^t \simeq \left(
\ba{ccc}
\cdot & -m_1 \cos \theta_\oplus & m_1 \sin \theta_\oplus \\
-m_1 \cos \theta_\oplus & \cdot & \cdot \\
m_1 \sin \theta_\oplus & \cdot & \cdot
\ea
\right) \quad .
\ee
In our 6D model, this pattern does not appear automatically if operators in Eq.~(\ref{eq:LD})
all have order one dimensionless coefficients. However, as we demonstrate in this numerical example,
it can appear if some operators are dominant. Moreover, this example is also interesting as it only requires 
operators with the lowest dimension in each sector.

As in the first numerical example, we take  $\theta_\Phi = 0.07$, $\theta_A = 1$, $\theta_\psi = \pi$, 
$\Lambda=10^3$~TeV and $M=1/R=70$~TeV. Let us start with the neutrino see-saw matrix. 
If we take $\tilde{Y}_\nu^+=y_\nu \, \{1,1.7\}$ for $s_+=\{-1,0\}$, $\tilde{Y}_\nu^-=y_\nu$ for $s_-=1$
with $y_\nu = 2.8 \cdot 10^{-2}$ and neglect all other operators, 
then the neutrino see-saw matrix is given by
\be
M_\nu = \left(
\ba{ccc}
0 & 3.62 \cdot 10^{-2} & 3.50 \cdot 10^{-2} \\
3.62 \cdot 10^{-2} & 1.46 \cdot 10^{-3} & 0 \\
3.50 \cdot 10^{-2} & 0 & 0
\ea
\right) \quad {\rm [eV]} \quad .
\ee
$M_\nu$ is diagonalized by one unitary matrix $U_\nu$
\be
U_\nu^\dagger M_\nu U_\nu^* = D_\nu =
{\rm diag}\{-5.003 \cdot 10^{-2}, 5.079 \cdot 10^{-2}, 7.089 \cdot 10^{-4} \} \quad {\rm [eV]} \quad ,
\ee
with
\be
U_\nu = \left(
\ba{ccc}
0.710 & 0.704 & 0.014 \\
-0.499 & 0.517 & -0.695 \\
-0.497 & 0.486 & 0.718
\ea
\right) \quad .
\ee
So $U_\nu$ has approximately the so-called bimaximal structure (see e.g. Ref.~\cite{Jezabek:1999ta} 
and references therein), which can easily be made compatible with the observed $U_{MNS}$, 
after small angles present in $U_l$ come into play.
The neutrino mass spectrum is pseudo-Dirac, with $\Delta m^2_{13} = 2.50 \times 10^{-3}~{\rm eV^2}$, 
$\Delta m^2_{21} = 7.63 \times 10^{-5}~{\rm eV^2}$, and $\Delta m^2_{21}/\Delta m^2_{13} = 3.05 \%$.

For the charged lepton matrix, we take a simple situation where only the two operators corresponding
to $S^l_-=\{X,\Phi\}$ are present with a dimensionless coefficient $y_0=7.0$. We get
\be
M_l = \left(
\ba{ccc}
4.21 \cdot 10^{-4} & 1.08 \cdot 10^{-3} & 0 \\
0 & 4.19 \cdot 10^{-3} & 5.98 \cdot 10^{-2} \\
0 & 0 & 1.71
\ea
\right) \quad [{\rm GeV}] \quad .
\ee
This matrix leads to hierarchical charged lepton masses, although somewhat discrepant with the
known values of $m_e$, $m_\mu$ and $m_\tau$. 
Here we need to stress again that the present evaluations are illustrative,
and use rough approximations of the fermion field profiles (Eqs.~(\ref{eq:fermprof})) and
step-up or linear profiles for the scalar fields (Eqs.~(\ref{eq:scalprof})).
So we have here
\be
U_l^\dagger M_l V_l = D_l = {\rm diag}\{4.07 \cdot 10^{-4}, 4.33 \cdot 10^{-3}, 1.71 \} \quad {\rm [GeV]} \quad ,
\ee
with
\be
U_l = \left(
\ba{ccc}
0.967 & 0.253 & 0.0 \\
-0.253 & 0.967 & 0.035  \\
0.009 & -0.034 & 0.999
\ea
\right) \quad , \quad
V_l = \left(
\ba{ccc}
0.999 & 0.025 & 0.0 \\
-0.025 & 0.999 & 0.0  \\
0.0 & 0.0 & 1
\ea
\right) \quad .
\ee
In this case, the mixing $e-\mu$ is responsible for the non-maximal mixing angle observed in solar neutrino
oscillations. We have for $U_{MNS} = U_l^\dagger U_\nu$
\be
U_{MNS} = \left(
\ba{ccc}
0.808 & 0.555 & 0.196 \\
-0.286 & 0.662 & -0.693 \\
-0.514 & 0.504 & 0.694
\ea
\right) \quad ,
\ee
corresponding to $\tan^2 \theta_{12} = 0.471$, $\tan^2 \theta_{23} = 0.997$, 
and $\sin^2 \theta_{13} = 3.85 \cdot 10^{-2}$.

Again, the pseudo-Dirac structure leads to a partial suppression of the effective neutrino Majorana mass
relevant for neutrinoless double-beta decay experiments
\be
|\left< m_{\beta \beta} \right>| = 17.0 \quad {\rm meV} \quad .
\ee
%

\section{Summary \& Conclusions}
\label{sec:end}

We returned to the question of neutrino masses in the context of a six-dimensional model compactified on a sphere.
Previously, we have dealt with charged fermions in this context, and shown how a vortex with winding number 3
allowed to generate 3 light 4-dimensional families from a single one in 6D.
The scheme is furthermore quite predictive, and mass hierarchies appear automatically.
We also noted that higher excitations of the gauge bosons mediate interesting neutral flavor-changing,
but family-number conserving interactions.
In fact, the winding number in the extra dimensions acts as a family number.

Here, we addressed specifically neutrinos (which we had only considered this far in a flat geometry),
and showed that, in addition to treating neutrinos like the other fermions (which results in Dirac masses),
light masses can be generated via the seesaw mechanism with the introduction of a single heavy neutrino in 6D, 
unbound to the vortex, embedded with a bulk Majorana mass.
As shown in Sec.~\ref{sec:majorana}, a distinctive feature of a Majorana mass in 6D is that it only
connects degrees of freedom (4D chiral components of the 6D field) with opposite 6D chiralities ``+" and ``-".
This feature combined with the particular structure of the fermion chiral zero modes in the vortex background,
for which the ``+" and ``-" components are explicitly dependent on each other (Eq.~(\ref{eq:Ln})), results
in a light neutrino mass matrix where one mixing angle is automatically maximal and where the eigenvalues
obey an inverted hierarchy with a pseudo-Dirac pattern for the heavier states $m_1 \simeq -m_2 \gg m_3$.

In this paper, the vortex paradigm is specifically implemented with a compactification on a sphere, 
which enables to consider bulk modes of the unbound field in both a consistent and calculable way. 
The size of the extra-dimensions is only constrained by limits on flavour violating processes, with the 
main constraint stemming from the decay $K \rightarrow \mu^\pm e^\mp$, which requires $1/R \geq 64$~TeV.
The size of the vortex on the other hand, which is here parameterized by the quantity 
$\delta = \theta_\Phi/\theta_A \sim 0.1$, and which governs the wave function profiles of the chiral zero-modes,
is chosen to match the steep hierarchies found in quark and charged lepton masses.
With this rather large value of $\delta$, the overlap between the zero-modes of the lepton doublet field $L$ 
and the singlet field $N$ is not suppressed enough to account for the smallness of the light neutrino masses.
For $1/R \sim 100$~TeV, these require a large bulk Majorana mass $M \sim 10^{11}$~GeV or,
a small Majorana mass $M \sim 10$~GeV or, more elegantly, $M \sim 1/R$ with slightly smaller dimensionless 
Yukawa couplings $\sim 10^{-3}-10^{-2}$ (one could also consider much smaller extra-dimensions $1/R \sim 10^6$~TeV).
While the model does not fix automatically the absolute neutrino mass scale, a very crucial point is that
it does fix the ratio of the mass squared differences, as $\Delta m^2_{21}/\Delta m^2_{31} \sim \delta^2$.
As we now know, the central value for the ratio of the observed $\Delta m^2$ for solar and atmospheric 
neutrinos is about 3.2\% in the inverted hierarchy scheme, therefore this value gives a strong support to this model.
Another strong clue in favor of the model is that the so-called solar mixing angle is predicted in the range 
$\pi/4 - \theta_\odot \sim \delta$, in very good agreement with the observed value $\theta_\odot \simeq 0.6$.

As shown in numerical examples 1 and 2 (Sec.~\ref{sec:ex1} and Sec.~\ref{sec:ex2}), realistic 
4D patterns for neutrino masses and mixings are possible in this model.
The main challenge is to account for the second large (and close to maximal) mixing angle first 
observed in atmospheric neutrino oscillation experiments.
To this end, some tuning of the dimensionless couplings in the 6D theory is necessary.
However, tuning does not necessarily mean fine-tuning. For instance, in the numerical example 2, the 
bimaximal structure in $U_\nu$ comes out naturally with all the non-zero coefficients in the neutrino sector 
having the same order of magnitude. The situation is quite different in the numerical example 1,
when one attempts to attribute the large atmospheric mixing angle to the charged lepton sector in $U_l$.
Although possible, this requires a hierarchy in the dimensionless coefficients 
$\tilde{Y}_l^-(0)/\tilde{Y}_l^-(1) \simeq \delta/2$, which breaks the simple order of magnitude reasoning.
Moreover, the coefficient $\tilde{Y}_l^-(-1)$ needs also to be adjusted quite finely if
one wants to keep the steep charged lepton mass hierarchy, in particular the smallness of the electron mass. 

Finally, this model (where we have not yet attempted to include possible CP violation) gives a number of 
definite predictions that are experimentally testable, but that have not been experimentally decided yet.
The first one is that the neutrino mass hierarchy is inverted, and so $\Delta m^2_{31} < 0$.
The second one is that the yet unknown small mixing angle $U_{e3}$ should be of order $\delta \sim 0.1$.
The third one emerges from the pseudo-Dirac pattern for the heaviest mass eigenstates $m_1 \simeq -m_2$,
which implies a partially suppressed signal at neutrinoless double-beta decay experiments.
We have $|\left< m_{\beta \beta} \right>| = |\sum_i m_i U_{e i}^2| 
\simeq \sqrt{\Delta m^2_{13}} \left( \frac{2}{3}-\frac{1}{3} \right) \simeq 16$~meV.

\section*{Notations}

Dirac fermions in six dimensions are described by eight-component spinors;
we work with the following representation of six-dimensional $8 \times 8$
Dirac matrices $\Gamma^A$ ($A = 0 \dots 5$)
\be
\Gamma^A = \left(
\ba{cc}
0 & \Sigma^A \\
\bar{\Sigma}^A & 0
\ea
\right) \quad ,
\ee
where $\Sigma^0 = \bar{\Sigma}^0 = \gamma^0 \gamma^0$,
$\Sigma^i = -\bar{\Sigma}^i = \gamma^0 \gamma^i$ ($i = 1 \dots 3$),
$\Sigma^4 = -\bar{\Sigma}^4 = i \gamma^0 \gamma^5$,
$\Sigma^5 = -\bar{\Sigma}^5 = \gamma^0$,
and $\gamma^\mu$ ($\mu = 0 \dots 3$), $\gamma^5 = i \gamma_0 \dots \gamma_3$
are the usual four-dimensional Dirac matrices in the chiral representation.
Also, the matrix $\Gamma_7$ is introduced as an analog of the matrix $\gamma_5$
in four dimensions
\be
\Gamma_7 = \Gamma^0 \dots \Gamma^5 =
\left(
\ba{cc}
1 & 0 \\
0 & -1
\ea
\right) \quad .
\ee
The subset of $\Gamma^\mu$ with $\mu = 0 \dots 3$ forms a $8 \times 8$ representation of the Clifford algebra in 4D.
Therefore, chirality in 4D is represented by the matrix $\tilde{\Gamma}_5 = -i \Gamma^0 \dots \Gamma^3$.
Finally, we define the matrix $\Sigma = i \Gamma^4 \Gamma^5$, therefore $\Gamma_7 = \tilde{\Gamma}_5 \Sigma$.

The dimensional couplings in the 6D theory are labeled by a capital $Y$ (\emph{e.g.} $Y_\nu^\pm$, $Y_l^\pm$).
The dimensionless couplings are denoted by the correspondent symbol with a tilde 
($\tilde{Y}_\nu^\pm$, $\tilde{Y}_l^\pm$). 
We use the symbols $M$, $M_D$, $M_\nu$ and $M_l$ for the neutrino Majorana mass in 6D,
the 4D neutrino Dirac masses, the light neutrino see-saw mass matrix, and the charged lepton mass matrix
respectively. Also, $n,m=1,2,3$ label the three families (generations) in the Standard Model.

\acknowledgments

The authors are indebted to E.~Nugaev and S.~Troitsky for useful comments
and discussions. This work is funded in part by IISN and by Belgian
Science Policy (IAP VI/11). This work has been supported in part (ML) by
the Russian Foundation for Basic Research grant 08-02-00473, by the
Federal Agency for Science and Innovations under state contract
02.740.11.0244, and by the grant of the President of the Russian
Federation NS-5525.2010.2. JMF thanks the CERN division for its hospitality in early 2009 while
this work was in progress. ML would like to thank the Service de
Physique Th\'{e}orique, at Universit\'{e} Libre de Bruxelles and Yukawa
Institute for Theoretical Physics, at Kyoto University where part of this
work was done under support in part by the Dynasty Foundation, for kind
hospitality. FSL would like to thank the Institute for Particle Physics
Phenomenology at Durham University where part of this work was done  for
kind hospitality.

\bibliographystyle{JHEP}
\bibliography{vortexnu}

\end{document}